\documentclass[aps,prl,preprint,superscriptaddress]{revtex4-1}
\usepackage{hyperref}
\usepackage{bm,amsmath}
\usepackage{graphicx,psfrag}
\usepackage{color}
\usepackage{amsmath, latexsym, amsfonts, amssymb, amsthm, amscd}
\newcommand \be  {\begin{equation}}  
\newcommand \bea {\begin{eqnarray} \nonumber }  
\newcommand \ee  {\end{equation}}  
\newcommand \eea {\end{eqnarray}}  

\begin{document}

\title{Quantum Mechanics with a non-zero quantum correlation time}
\author{Jean-Philippe Bouchaud}
\affiliation{
Capital Fund Management, 23 rue de l'Universit\'e, 75007 Paris, France.}

\date{\today}

\begin{abstract}
We propose an extension of Quantum Mechanics based on the idea that the underlying ``quantum noise'' has a non-zero, albeit very small, correlation time $\tau_c$. 
The standard (non-relativistic) Schrodinger equation is recovered to zeroth order in $\tau_c$, and the first correction to energy levels is computed. Some consequences are discussed, 
in particular the violation of Heisenberg's uncertainty principle and the restoration of locality at short times. 
\end{abstract}


\maketitle

The existence of a dimensional fundamental constant generally leads to new physical phenomena, with Planck's black body radiation law ($\hbar$) or Einstein's special 
relativity ($c$) as canonical examples. Here we want to explore the possibility that the ``quantum noise'' underlying Feynman's random paths view of quantum mechanics is actually not 
a white noise but has some very small but non-zero correlation time $\tau_c$. We derive an extension of Schrodinger's equation to the case where the noise is an Ornstein-Uhlenbeck 
process, and show how the standard (non-relativistic) Schrodinger equation is modified to first order in $\tau_c$. Although at this point not much more than an intellectual exercise, 
leading to an interesting problem for an advanced quantum mechanics course, the existence of a non-zero quantum correlation time is related to more fundamental questions currently debated in the literature.

We start by recalling Feynman's view of the quantum evolution of a particle as a classical Brownian motion in imaginary time \cite{FH}. In order to keep notations simple, we restrict to 
the one dimensional motion of a non relativistic particle of mass $m$ in a potential $V(x)$. The quantum probability amplitude $\psi(x,t)$ can written as:
\be
\psi(x,t) = \int_{\text{paths}} e^{{\rm i} \frac{S[\text{path}]}{\hbar}} \psi(x_0,0); \qquad S[\text{path}] = \int_0^t {\rm d}s \left[\frac{m}{2} \left(\frac{dx}{ds}\right)^2 - V(x(s))\right],
\ee
from which, as is well known, one can derive the usual Schrodinger equation using standard manipulations \cite{FH}:
\be
{\rm i} \hbar \frac{\partial \psi}{\partial t} = - \frac{\hbar^2}{2m} \nabla_x^2 \psi + V(x) \psi.
\ee
The kinetic energy term in the action $S$ is tantamount to assuming that the particle's velocity is a Gaussian white noise (in imaginary time), i.e.:
\be
m \frac{dx}{dt} = p(t); \qquad \langle \, p(s) p(u) \, \rangle = m \hbar \, \delta(s-u).
\ee
The main idea of the present paper is to relax the assumption of a zero quantum correlation time and consider that $p(t)$ is an Ornstein-Uhlenbeck process, i.e. we posit that:
\be
\langle \, p(s) p(u) \, \rangle = \frac{m \hbar}{2 \tau_c} \exp - \frac{|s-u|}{\tau_c}.
\ee
Since the operator inverse of the exponential correlation function is $\propto \delta(s-u) \left[1 + \tau_c^2 \partial_s \partial_u\right]$, the associated path integral 
representation evolution of the probability amplitude for $x$ {{\rm i}t and} $p$ reads:
\be\label{eq:gen_action}
\Psi(x,p,t) = \int_{\text{paths}} e^{{\rm i} \frac{S[\text{path}]}{\hbar}} \Psi(x_0,p_0,0); \quad S[\text{path}] = \int_0^t {\rm d}s \left[\frac{1}{2 m}\left(p^2 - \tau_c^2 (\frac{dp}{ds})^2\right) 
- V(x(s))\right],
\ee
where the path integral is restricted to paths where $dx/ds = p(s)$ for all $s \in [0,t]$.  
Alternatively, our assumption of an Ornstein-Uhlenbeck correlator can be seen as retaining the first non-trivial 
term involving time derivatives of $p$ in the action $S$. From the above path integral representation one can derive a generalized Schrodinger equation that reads:
\be
{\rm i} \hbar \frac{\partial \Psi(x,q)}{\partial t} = \frac{\hbar}{2 \tau_c} \left[  \nabla^2_q \Psi(x,q) - q^2 \Psi(x,q)\right] + {\rm i} \sqrt{\frac{\hbar^3}{m \tau_c}} \, q \nabla_x \Psi(x,q) + {V(x)} \Psi(x,q),
\ee
where we have rescaled $p$ as $p = \sqrt{{m \hbar}/{\tau_c}} q$. One can check that this extended Schrodinger evolution equation is unitary, i.e. it 
conserves the total norm of $\Psi$, $\iint {\rm d}x{\rm d}q |\Psi(x,q)|^2=1$, $\forall t$. 

Looking for a time-stationary solution of the form $\Psi \propto e^{-{\rm i} Et/\hbar}$, one finds the following generalized eigenvalue equation, which is the central 
result of this paper:
\be\label{eq:Schr}
E \Psi(x,q) =  \frac{\hbar}{2 \tau_c} \left[\nabla^2_q \Psi(x,q) - q^2 \Psi(x,q) \right]  + {\rm i} \sqrt{\frac{\hbar^3}{m \tau_c}} \, q \nabla_x \Psi(x,q) + V(x) \Psi(x,q).
\ee
In view of the harmonic oscillator form of the operator in $q$, it is natural to look for solutions of the form:
\be\label{expand}
\Psi(x,q) = \sum_{n=0}^\infty \xi_n(x) \mathcal{H}_n(q),
\ee
where $\mathcal{H}_n(q)$ are the standard harmonic oscillator eigenstates, indexed by $n$. The probability of presence of the particle at position $x$ is then obtained as 
$\mathbb{P}(x)= \int {\rm d}q |\Psi(x,q)|^2 = \sum_{n=0}^\infty |\xi_n(x)|^2$, where we have used the orthogonality of the ${\cal H}_n(q)$. 

Multiplying Eq. (\ref{eq:Schr}) by $\mathcal{H}_n(q)$ and integrating over $q$ leads to:
\be\label{eq:Schr2}
E \xi_n(x) = - \frac{\hbar}{\tau_c} (n + \frac12) \xi_n(x) + V(x) \xi_n(x) +  {\rm i} \sqrt{\frac{\hbar^3}{2m \tau_c}} \left[\sqrt{n} \xi_{n-1}'(x) + \sqrt{n+1} \xi_{n+1}'(x)\right],
\ee
where we have used the well-known result
\be\label{eq:Hn}
q \mathcal{H}_n(q) = \sqrt{\frac{n+1}{2}} \mathcal{H}_{n+1}(q) + \sqrt{\frac{n}{2}} \mathcal{H}_{n-1}(q).
\ee

In analogy with similar calculations in the context of the Fokker-Planck equation \cite{Gardiner}, we will now analyze Eq. \ref{eq:Schr2} perturbatively in $\tau_c$. 
We make the following ansatz:
\be
E = \frac{\mathcal{E}_0}{\tau_c} + \mathcal{E}_1 + \tau_c \mathcal{E}_2 + \dots
\ee
and
\be
\xi_n(x)= (\tau_c)^{n/2} \left[\phi_{n,0}(x) + \tau_c \phi_{n,1}(x) + \dots \right].
\ee
Plugging these expressions into Eq. (\ref{eq:Schr2}) for $n=0$, we get, to leading order in $\tau_c$ (we drop the explicit $x$ dependence henceforth):
\be
\frac{\mathcal{E}_0}{\tau_c} \xi_0 + \mathcal{E}_1 \phi_{0,0} + \tau_c (\mathcal{E}_1 \phi_{0,1}+\mathcal{E}_2 \phi_{0,0}) = - \frac{\hbar}{2\tau_c} \xi_0 + V (\phi_{0,0} + \tau_c \phi_{0,1})
+  {\rm i} \sqrt{\frac{\hbar^3}{2m}} (\phi'_{1,0} + \tau_c \phi'_{1,1});
\ee
from which one obtains:
\be
\mathcal{E}_0 = -\frac{\hbar}{2},
\ee
\be\label{eq:1}
\mathcal{E}_1 \phi_{0,0} = V \phi_{0,0} +  {\rm i} \sqrt{\frac{\hbar^3}{2m}} \phi'_{1,0},
\ee
and
\be\label{eq:2}
\mathcal{E}_1 \phi_{0,1}+\mathcal{E}_2 \phi_{0,0} = V \phi_{0,1} +  {\rm i} \sqrt{\frac{\hbar^3}{2m}} \phi'_{1,1}.
\ee
Similarly, Eq. (\ref{eq:Schr2}) for $n=1$ yields:
\be\label{eq:3}
\phi_{1,0} = {\rm i} \sqrt{\frac{\hbar}{2m}} \phi'_{0,0}
\ee
and
\be\label{eq:4}
\mathcal{E}_1 \phi_{1,0} = - \frac{\hbar}{2} \phi_{1,1} + V \phi_{1,0} + {\rm i} \sqrt{\frac{\hbar^3}{2m}}  (\phi'_{0,1} + \sqrt{2} \phi'_{2,0}).
\ee
Finally, the leading term (or order $\tau_c^0$) for $n=2$ gives:
\be\label{eq:5}
\phi_{2,0} =  \frac{{\rm i}}{2} \sqrt{\frac{\hbar}{m}} \phi'_{1,0}.
\ee

Now, inserting Eq. (\ref{eq:3}) into Eq. (\ref{eq:1}) recovers to the usual Schrodinger equation for $\phi_{0,0}$:
\be\label{eq:Sch_again}
\mathbb{H}_m \phi_{0,0} = \mathcal{E}_1 \phi_{0,0}, \qquad  \mathbb{H}_m := - {\frac{\hbar^2}{2m}} \nabla_x^2 + V(x),
\ee
which is the limiting equation in the limit $\tau_c \to 0$, as it should be. Quite remarkably, one finds exactly the same Schrodinger equation if one 
assumes that $q$ is not in the ground state $n=0$ but {\it in any other state} $n$: the dependence on $n$ cancels out entirely, as a consequence of Equation \eqref{eq:Hn}.

The next order correction is obtained by first combining Eq. (\ref{eq:4}) with 
Eqs. (\ref{eq:3},\ref{eq:5}) to give:
\be
\phi'_{1,1} =  {\rm i} \sqrt{\frac{2}{m \hbar}} \left[(V-\mathcal{E}_1) \phi'_{0,0}\right]'  + {\rm i} \sqrt{\frac{2\hbar}{m}} \phi''_{0,1} - {\rm i} \sqrt{\frac{\hbar^3}{2m^3}}  \phi^{(4)}_{0,0},
\ee
where the last term is the fourth derivative of $\phi_{0,0}$. Inserting this last expression into Eq. (\ref{eq:2}) finally leads to an equation for the first correction to the 
$n=0$ sector wave-function $\phi_{0,1}$:
\be
(V -\mathcal{E}_1) \phi_{0,1} - {\frac{\hbar^2}{m}} \phi''_{0,1} =  \mathcal{E}_2 \phi_{0,0} + {\frac{\hbar}{m}} \left[(V-\mathcal{E}_1) \phi'_{0,0}\right]' - \frac{\hbar^3}{2m^2} \phi^{(4)}_{0,0}.
\ee
Using Eq. (\ref{eq:Sch_again}) to eliminate the fourth derivative term, this can be rewritten as:
\be
(V -\mathcal{E}_1) \phi_{0,1} - {\frac{\hbar^2}{m}} \phi''_{0,1} =  (\mathcal{E}_2 - \frac{\hbar}{m} V'') \phi_{0,0} -  \frac{\hbar}{m} V' \phi'_{0,0}.
\ee
Let us denote as $\Phi_{\varepsilon}(x)$ the eigenstates of $\mathbb{H}_m$ corresponding to energies $\mathcal{E}_1$, and $\mathbb{H}_{m/2}$ the Hamiltonian corresponding 
to a particle of mass $m/2$ in the same potential $V(x)$. The last equation is formally solved as (using the bracket notation):
\be
\vert \phi_{0,1} \rangle = (\mathbb{H}_{m/2} - \mathcal{E}_1 \mathbb{I})^{-1}(\mathcal{E}_2 - \frac{\hbar}{m} V''  - \frac{\hbar}{m} V' \partial_x) \vert \Phi_{\varepsilon} \rangle.
\ee
This last equation can be used to compute the first correction $\mathcal{E}_2 \tau_c$ to the Schrodinger energy levels $\mathcal{E}_1$, induced by virtual transitions between the $n=0$ and $n=1$ levels in the $p$-sector.
This correction is obtained by noting that for the consistency of our expansion in $\tau_c$, the correction $\phi_{0,1}$ must be such that $\langle \phi_{0,0} \vert \phi_{0,1} \rangle=0$, leading to:
\be
\mathcal{E}_2 \tau_c= \frac{\hbar  \tau_c}{m} \frac{\langle \Phi_{\varepsilon} \vert (\mathbb{H}_{m/2} - \mathcal{E}_1 \mathbb{I})^{-1}(V'' + V' \partial_x) \vert \Phi_{\varepsilon} \rangle}
{\langle \Phi_{\varepsilon} \vert (\mathbb{H}_{m/2} - \mathcal{E}_1 \mathbb{I})^{-1} \vert \Phi_{\varepsilon} \rangle},
\ee
This expression is completely general and gives the first correction in $\tau_c$ to the energy levels for an arbitrary (one-dimensional) quantum mechanical problem. 
Note that this correction is zero for a free particle (i.e. $V' \equiv 0$); the familiar $\mathcal{E}_1(k) = \hbar^2 k^2/2m$ is in fact exact to all orders in $\tau_c$. 
Another fully soluble case to all orders in $\tau_c$ is the harmonic oscillator, $V(x)=m \omega^2 x^2/2$, in which case Eq. \eqref{eq:Schr} is known as the Pais-Uhlenbeck oscillator \cite{Pais}.
In this case, the energy levels $E_{n,\ell}$ are indexed by two integers $n, \ell$:
\be
E_{n,\ell} = -  \frac{\hbar}{\tau_c} \sqrt{1 - (\omega \tau_c)^2} (n + \frac12) + \hbar \omega (\ell + \frac12).
\ee
For a fixed $n$, one recovers the familiar harmonic oscillator levels, all shifted by an amount $(2n+1)\hbar \omega^2 \tau_c/4$ for $\omega \tau_c \ll 1$. For $\omega > 1/\tau_c$, the problem becomes ill-defined, presumably 
because new physics sets in when the frequency of the harmonic oscillator becomes comparable to the quantum noise correlation time.

An interesting consequence of a non-zero quantum correlation time concerns Heisenberg's uncertaintly principle. As shown by Feynman \cite{FH2}, the $\tau_c=0$, Brownian noise limit can be formally associated to the 
non-commutation of the position $X$ and momentum $P$ operators. One can extend Feynman's argument \cite{FH2} to $\tau_c \neq 0$ for (time) coarse-grained operators:
\be
\widehat X_\tau = \frac{1}{\tau} \int_{t-\tau/2}^{t+\tau/2} {\rm d}t' \, X(t'); \qquad \widehat P_\tau = \frac{1}{\tau} \int_{t-\tau/2}^{t+\tau/2} {\rm d}t' \, P(t'),
\ee
with the result:
\be
\left[\widehat X_\tau ,\widehat P_\tau\right] =\i \hbar \left(1 - \frac{(1-e^{-u})^2}{u^2}\right); \qquad u:=\frac{\tau}{\tau_c}.
\ee
Hence, the standard uncertainty relation is recovered in the limit $\tau \gg \tau_c$ (with a negative correction of order $(\tau_c/\tau)^2$), but is violated for $\tau \sim \tau_c$ (see Fig. 1).
It would be interesting to see whether presently available experiments can be used to bound $\tau_c$ from above, but this would require extending the
above calculations to quantum fields. Another direction is to compute tunnelling amplitudes that should be highly sensitive to $\tau_c$ is the high barrier, low transmission limit. 

\begin{figure}[t]
\includegraphics[width=8cm, height=6cm,clip=true]{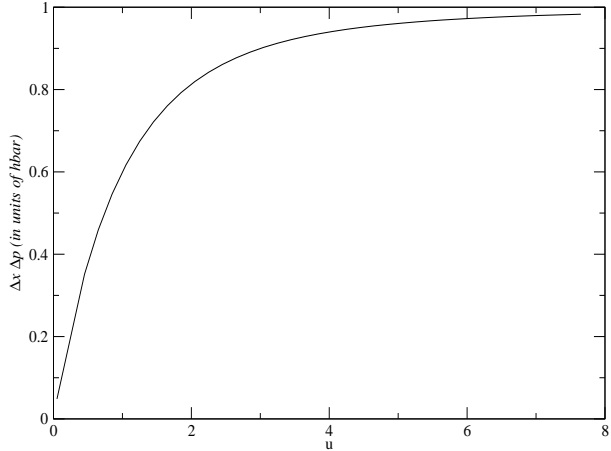}
\caption{\small Violation of Heisenberg's uncertainty principle. We plot the minimum value of $\Delta x \Delta p/\hbar$ as a function of the ratio $u:=\tau/\tau_c$, where $\tau$ is the measurement time. 
The uncertainty vanishes linearly when $u \to 0$, and saturates to the standard $\hbar$ value for $\tau \gg \tau_c$, with a correction going down as $u^{-2}$.}
\label{fig1}
\end{figure}

Although enticing from a purely theoretical point of view, it is not clear whether the above generalisation of quantum mechanics has any relevance to describe the real world. However, from a purely 
logical point of view, there is no reason why the quantum correlation time should be zero, or said differently why quantum mechanical actions should not contain terms with higher order time 
derivatives, as in Eq. (\ref{eq:gen_action}). This has been argued by various authors, starting with Pais and Uhlenbeck \cite{Pais, Hawking}, see also \cite{Baulieu} in another context. 
One crucial point, however, is to interpret the unbounded negative energy spectrum coming from the $p$-sector, that correspond to the classical Ostrogadsky theorem for higher-order time derivatives in the Hamiltonian \cite{Ostro}. 
There has been a flurry of papers on that subject recently, see e.g. \cite{Mannheim,Bender,Woodard,review}. Physically, unbounded negative energies are only problematic if real transitions towards 
$n \geq 1$ states in the $p$ sector are possible. This would require strange terms in the Hamiltonian coupling $x$ and $p$ allowed by symmetry, such as $x^2 \times p^2$, or some coupling to a yet unknown field. 
Intriguingly, the Schrodinger limit is recovered for any level $n$ level and its virtual excitations to the $n+1$ and $n-1$ levels. 
This invariance may play an important role for the viability of the theory, since the low energy physics is blind to any transition taking place in the $p$-sector.

I thank L. Baulieu, R. B\'enichou, B. Chowdury, J. Dalibard, T. Gueudr\'e \& A. Zee for insightful remarks. I want to warmly thank Antoine Tilloy and Zura Kakushadze for pointing out the link 
with the Ostrogradsky problem and Pais-Uhlenbeck oscillator.

\end{document}